\documentclass{article}

\usepackage{amsmath,amsthm,amssymb,url}
\usepackage{epsfig}

\usepackage{fullpage}

\newtheorem{theorem}{Theorem}[section]

\newtheorem{corollary}[theorem]{Corollary}
\newtheorem{proposition}[theorem]{Proposition}
\theoremstyle{definition}

\newtheorem{example}[theorem]{Example}

\newcommand{\xa}{\star}
\newcommand{\xb}{\Box}
\newcommand{\xc}{\triangle}
\newcommand{\xt}{\times}
\newcommand{\xd}{\nabla}

\begin{document}

\title{Outer bounds for exact repair codes}

\author{Iwan M. Duursma}

\date{\today}

\maketitle

\begin{abstract}
We address the open problem of establishing the rate region for exact-repair regenerating codes for given parameters $(n,k,d)$. Tian determined the rate region for a $(4,3,3)$ code and found that it lies strictly within the functional-repair rate region.
% obtained by Dimakis, Godfrey, Wu, Wainwright and Ramchandran. 
Using different methods, Sasidharan, Senthoor and Kumar prove a non-vanishing gap between the functional-repair outer bound and the exact-repair outer bound for codes with $k \geq 3$. Our main results are two improved outer bounds for exact-repair regenerating codes. They capture and then extend essential parts in the proofs by Tian and by Sasidharan, Senthoor and Kumar. We show that the bounds can be combined for further improvements.
\end{abstract}

%An open problem is to establish the rate region for exact-repair regenerating codes with given parameters $(n,k,d)$. Tian determined the rate region for the $(4,3,3)$ code and found that it lies strictly within the functional-repair rate region.
% obtained by Dimakis, Godfrey, Wu, Wainwright and Ramchandran. 
%Using different methods, Sasidharan, Senthoor and Kumar prove a non-vanishing gap between the exact-repair outer bound and the functional-repair outer bound for codes with $k \geq 3$. Our main results are two different outer bounds for exact-repair regenerating codes. %They capture essential parts in the proofs other extends a crucial part in the proof of Tian. One extends the methods of Sasidharan, Senthoor and Kumar. The other extends a crucial part in the proofof Tian. When used separately the two bounds improve known results, with %further improvements when used in combination.
%\end{abstract}

\section*{Introduction}
Regenerating codes were introduced by Dimakis, Godfrey, Wu, Wainwright and Ramchandran \cite{DGWWR10}. Their main application is in large distributed storage systems where they lead to significant savings by optimizing the trade-off between storage size and repair bandwith. In a distributed storage system (DSS) an encoded file is stored on $n$ servers such that it can be recovered from any combination of $k$ servers. If a server fails it can be rebuilt by retrieving the information needed for its repair from any combination of $d$ other servers. An encoding scheme realizing these parameters is called an $(n,k,d)$ regenerating code. For background and details on distributed storage and regenerating codes we refer to \cite{DRWS11}, \cite{SRKR10}. A common example is the use of a $(4,2,3)$ code to store four bits $x,y,z,t$.
By storing the pairs of bits $(x,z+t), (y,t+x), (z,x+y), (t,y+z)$ on four different servers ($n=4$), the four bits $x,y,z,t$ can be recovered from the combined information on any two servers ($k=2$). And if a server fails it can be rebuilt by retrieving one bit from each of the remaining three servers ($d=3$). In particular, the first server can be rebuilt from the three bits $y, y+x,$ and $y+z+t$. 

An $(n,k,d)$ code comes with a secondary set of parameters $(B,\alpha,\beta)$. For a file of size $B$, a part of size at most $\alpha$ is stored on a single server, and bandwith between a server and any of the $d$ servers helping in its repair is limited to $\beta$. For the example, $B=4, \alpha=2, \beta=1$. The gains in a DSS are obtained by using a total repair bandwith $\gamma = d \beta$ that is possibly larger than $\alpha$ but much smaller than the file size $B$. 
The challenge is, given $(n,k,d)$, to optimize the trade-off between the storage $\alpha$ per server and the repair bandwith $\beta$ between servers in order to store a file of size $B$. For given parameters $(n,k,d)$, the outer bound refers to the relation among the parameters $(B,\alpha,\beta)$. The outer bound can be interpreted as an upper bound on the file size $B$, for given $\alpha$ and $\beta$, or as a lower bound for $\alpha$ and $\beta$, for a given file size $B$. In the first case it is standard to scale to variables $B/\beta$ and $\alpha/\beta$, and in the second case to variables $\alpha/B$ and $\beta/B$.  

In this work, we establish new outer bounds for exact-repair regenerating codes. In the exact-repair scenario it is required that a server be rebuilt to its original form. The weaker requirement, known as functional repair, only requires that a server be rebuilt to a form that preserves the functionality of the DSS. Upper bounds for the file size under functional repair are piece-wise linear and take the form
\begin{equation} \label{Bfr}
B \leq B_q = q\alpha +  \binom{k-q}{2} \beta + (d+1-k)(k-q) \beta, \quad 0 \leq q \leq k.
\end{equation}
Details of the bound and motivation behind the linear functions $B_q$ are recalled in Section \ref{S:1}. For the values $(n,k,d)=(4,2,3)$, 
\[
B \leq \min ( B_2, B_1, B_0 ) =  \min ( 2 \alpha, \alpha+ 2\beta, 5 \beta ).
\]
For the four bit example with $(B,\alpha,\beta)=(4,2,1)$ the bound is sharp at both $B \leq B_1$ and $B \leq B_2$.
The vertex with $B = B_k = B_{k-1}$ minimizes $\alpha$ and is called the MSR point (for Minimum Storage Regenerating code). The vertex with $B = B_1 = B_0$ minimizes $\beta$ and is called the MBR point (for Minimum Bandwith Regenerating code).
Both these points are achieved by exact-repair regenerating codes using the general construction in \cite{RSK11}. Using the notion of information flow graph and then applying results from network coding, the main result of \cite{DGWWR10} shows that the bound (\ref{Bfr}) is sharp for regenerating codes under the functional-repair requirement. Clearly, exact-repair implies functional-repair, and the outer bound (\ref{Bfr}) applies to exact-repair regenerating codes

Tian \cite{T14} determined the rate region for a $(4,3,3)$ code and provided the first example of an exact-repair rate region that lies strictly within the functional-repair rate region. For a $(4,3,3)$ code
\begin{equation} \label{B433}
B \leq \min ( B_3, B_2, B_1, B_0 \}
 =  \min ( 3 \alpha, 2 \alpha + \beta, \alpha + 3 \beta, 6 \beta \}.
\end{equation}
%Figure \ref{} (a) illustrates the bound (\ref{B433}) and Figure \ref{} (b) illustrates the dual interpretation as a trade-off between %minimizing the storage $\alpha$ versus minimizing the repair bandwith $\beta$. 
The exact-repair region is describeded by adding to (\ref{B433}) the new inequality
\begin{equation} \label{new}
3B ~\leq~ 4 \alpha + 6 \beta~ ( = 2B_1+B_2-\beta ).
\end{equation}
%which yields an improvement in the range $3\beta/2 < \alpha < 3 \beta$ and makes $B \leq B_1$ redundant.
To prove that all points in the new region can be achieved it suffices, using a standard time sharing argument, that the vertices with $\alpha/\beta \in \{ 1,\;3/2,\;3 \}$ can be achieved. The first and the last are the MSR and the MBR point. An example achieving $B=8, \alpha = 3, \beta =2$ is provided in \cite{T14}. A different example is to encode eight bits $x_1, x_2, y_1, y_2, z_1, z_2, t_1, t_2$ as four triples
\[
\begin{array}{ccccc}
x1 & &x2 & &z1+t2 \\
y1  & & (y2) & &(t1+x2) \\
(z1) & &z2 & &(x1+y2) \\
(t1) & &(t2) & &y1+z2
\end{array}
\]
The repair information to rebuilt the first server is given in parentheses. 

As part of their results, Sasidharan, Senthoor and Kumar \cite{SSK13} obtain the same inequality (\ref{new}). Their main result however [Ibid., Theorem 1] is a non-vanishing gap between the  functional-repair outer bound and the exact-repair outer bound for all codes with $k \geq 3$. 

Our results include the following outer bound (Theorem \ref{thm-rs}).

\begin{quote}
\emph{For given $k,d$, let $q,r,s$ be positive integers with $q+r+s \leq k$. Let $V_1,V_2,\ldots,V_{n-2}$ be subsets of $\{ r+s+1,\ldots,d+1 \}$ of size $q_i = |V_i| \leq k-r-s$ with empty intersection
$V_1 \cap V_2 \cap \cdots \cap V_{n-2} = \emptyset$. Then}
\[
n B \leq B_q + \sum_{i=1}^{n-2} B_{q_i} + B_{r+s} - rs \beta.
\]
\end{quote}

For $k=2p, d=3p$, $q=q_1=q_2=r+s=p$, it follows that $B \leq B_p - (p^2-1) \beta / 16$. This difference is unbounded as $p$ goes to infinity. On the other hand, the non-vanishing gap in \cite{SSK13} remains bounded and is always less than $\beta$.

A second outer bound (Theorem \ref{thm-lm}) is obtained using a similar approach as in \cite{SSK13}. For both outer bounds we give examples that illustrate the improvements over known bounds. We also show that the bounds can be combined for further improvements.

The next section formulates the main problem. Section \ref{S:2} presents the main arguments and how they are used in two different proofs for the rate region of a $(4,3,3)$-code. Section \ref{S:4} proves Theorem \ref{thm-rs}. Section \ref{S:3} builds on the approach used in \cite{SSK13} and proves Theorem \ref{thm-lm}. It also contains a short proof for a non-vanishing gap of $\beta/6$ between the outer bounds for  functional-repair and exact-repair.  Section \ref{S:5} illustrates how the results of the two previous sections can be used in combination. 

\section{An optimization problem on random variables} \label{S:1}

By an exact-repair regenerating code of type $(n,k,d)$ with secondary parameters $(B,\alpha,\beta)$ we mean a collection of random variables $M$, $\{ W_j : 1 \leq j \leq n \}$ and $\{ S_i^j : 1 \leq i,j \leq n, i \neq j \}$ that satisfy several  entropy constraints. Let $W_J$ denote the joint distributions $W_J = (W_j : j \in J)$ and, for $j \not \in I$, let $S_I^j$ denote the joint distribution $S_I^j = ( S_i^j : i \in I).$ The entropy constraints are the following
\begin{align}
&H(M) = B. \label{en1} \\
&H(W_j) = \alpha,~~~H(W_j|M) = 0,~~~H(M|W_J) = 0~(|J|\geq k).  \label{en2} \\
&H(S_i^j) = \beta,~~~H(S_i^j|W_i) =0,~~~H(W_j|S_I^j) = 0~(|I| \geq d, j \not \in I). \label{en3}
\end{align}
The interpretation for a distributed storage system is that $M$ is the file to be stored, $W_i$ is the part of the encoded file that is stored on server $i$, and $S_i^j$ is the helper information provided by server $i$ to repair server $j$. Assuming uniform distributions for each of the variables, the conditions $H(M)=B, H(W_i)=\alpha$ and $H(S_i^j)=\beta$ describe the size of the underlying space for $M, W_i$ and $S_i^j$, respectively. The condition $H(W_i|M) = 0$ says that the information stored on server $i$ is completely determined by the file $M$, and similarly $H(S_i^j|W_i) =0$ says that helper information provided by server $i$ is completely determined by information stored on server $i$. Finally, the access condition $H(M|W_J) = 0~(|J|\geq k)$ says that the file can be recovered from information stored on any $k$ servers, and similarly $H(W_j|S_I^j) = 0~(|I| \geq d, j \not \in I)$ says that server $j$ can be rebuilt with helper information received from any $d$ remaining servers.
Clearly, for $|J| =k$,
\[
B = H(M) = H(M|W_J)+H(W_J) \leq 0 + \sum_{j \in J} H(W_j) = k \alpha.
\]
and thus $B \leq k \alpha.$ Moreover, for $0 \leq q \leq k$, let $\{1,2,\ldots,d+1\} = V' \cup V \cup U$ be a partition with $|V'|=q, |V|=k-q$ and $|U|=d+1-k$, and let
$S_V$ denote the joint distribution $S_V = ( S_i^j : i,j \in V, i > j ).$ Then 
\[
H(M | W_{V'} S_V S_U^V) = 0 
\]
and
\begin{align*}
B = H(M) &\leq~ H(W_{V'} S_V S_U^V) \leq H(W_{V'}) + H(S_V) + H(S_U^V) \\
&\leq~ q \alpha + \binom{k-q}{2} \beta + (d+1-k)(k-q) \beta~~~(=:B_q)
\end{align*}
The upper bound $B \leq \min \{ B_q : 0 \leq q \leq k \}$ applies to both the exact-repair and the functional-repair setting and is sharp for the latter \cite{DGWWR10}.
We will make use of the following.
\begin{align}
&B_{r+1}-B_r = \alpha - (d-r)\beta,  ~~~(0 \leq r \leq k-1)  \label{Br} \\
&B_{r+m}-B_{r+m-1}+B_r-B_{r_1}= (m-1) \beta. ~~~(1 \leq r \leq r+m \leq k) \label{Brm} 
\end{align}

\subsection{The case $(n,k,d)=(4,3,3)$}

The single argument that we are aware of to improve the bound
\[
B \leq \min \{ B_q : 0 \leq q \leq k \}
\]
is to sum multiple copies of the bound for different choices of $(V',V,U)$ and to exploit nonzero mutual information among variables in the different copies. 
For the $(4,3,3)$-code, one such sum of three copies is
\[
3B ~\leq~ H(W_1 S_3^2 S_4^2 S_4^3) + H(W_4 S_3^2 S_1^1 S_1^3) + H(W_2 W_3 S_1^4),
\]
which corresponds to $(V',V,U)=(\{1\}, \{2,3\},\{4\}),$ $(\{4\},\{2,3\},\{1\}),$ $ (\{2,3\},\{4\},\{1\}).$ After adding a term $H(S_3^2)$, we can regroup the variables (details are provided in Section\ref{S:4}).
\begin{align*}
3B + \beta &\leq~ H(W_1 S_3^2 S_4^2 S_4^3) + H(W_4 S_3^2 S_1^2 S_1^3) + H(W_2 W_3 S_1^4) + H(S_3^2)  \\
                  &\leq~ H(W_1 S_3^2 S_4^2) + H(W_4 S_3^2 S_1^2) + H(W_3 S_1^4) + H(W_2 S_3^2 S_4^3 S_1^3) 
\end{align*}
Finally, $H(W_2 S_3^2 S_4^3 S_1^3) = H(W_2 S_4^3 S_1^3)$ leads to a savings of $\beta$ in the sum of the original three copies and thus to (\ref{new}).
The same argument is at the core of the proof in \cite{T14}. %With the benefit of hindsight that the proof exists it is possible to use the same argument here for a different shorter proof.

A different choice for the sum of three copies is
\[
3B ~\leq~  H(W_1 S_3^2 S_4^2 S_4^3) + H(W_2 W_3 S_4^1) + H(W_2 W_3 W_4),
\]
which corresponds to $(V',V,U)=(\{1\},\{2,3\},\{4\}),$ $(\{2,3\},\{1\},\{4\}),$ $ (\{2,3,4\},\emptyset,\{1\}).$
First we regroup the variables to obtain (details are provided in Section\ref{S:3})
\[
3B ~\leq~ H(W_1 S_3^2) + H(W_2 W_3 S_4^2) + H(W_2 W_3 S_4^3) + H(S_4^1) + H(W_4). 
\]
Using $H(W_2 W_3 S_4^2) = H(W_3 S_4^2) + H( W_2 | W_3 S_4^2) \leq H(W_3 S_4^2) + H(S_1^2)$ and similarly $H(W_2 W_3 S_4^3) \leq H(W_2 S_4^3) + H(S_1^3)$  leads to (\ref{new}). The final reduction follows \cite[Equation (18)]{SSK13}.

A difference between the first choice and the second choice is that the first choice reduces the number of $S_i^j$ while the second choice reduces the number of $W_i$ at a cost of increasing the number of $S_ i^j$.

\subsection{Relation to secret sharing}

We briefly point out the connection between regenerating codes and secret sharing.
In secret sharing a sensitive message is distributed over several parties such that only qualified combinations of parties can reconstruct the message. For the distribution of a message $M$ using shares $W_j$, $j=1,2,\ldots,n,$ let
\begin{align*}
&H(M|W_J) = 0, ~\text{for $|J| \geq r$,} \\ %\qquad \text{and} \qquad  
&I(M;W_J) = 0, ~\text{for $|J| \leq t$}.
\end{align*}
The minimal choice for $r$ (resp. the maximal choice for $t$) is called the acceptance threshold (resp. the rejection threshold).
For the distribution of a file $M$ over servers $W_j$, $j=1,2,\ldots,n$ we use the conditions
\begin{align} 
H(M|W_J) = 0, \quad \text{if $|J| \geq k$.} \label{eq:t1} \\ 
I(M;W_J) \leq \alpha, \quad \text{if $|J| \leq 1$.} \label{eq:t2}
\end{align}
For a server $W_j$, let $S^j_1, \ldots, S^j_m$ be helper nodes. We add a second set of conditions
\begin{align} 
H(W_j|S^j_I) = 0, \quad \text{if $|I| \geq d$.} \label{eq:b1}\\
I(W_j;S^j_I) \leq\beta, \quad \text{if $|I| \leq 1$}. \label{eq:b2}
\end{align}
These conditions express that a DSS is similar to a two-layer secret sharing scheme where the condition on mutual information is relaxed from zero mutual infirmation to bounded mutual information. The condition of bounded mutual information is enforced to obtain efficient storage in the top-layer and efficient bandwith in the bottom layer (rather than to reduce the information about the secret as in an actual secret sharing scheme). The two-layered secret scheme becomes a regenerating code if we enforce that a share $S^j_i$ is stored not on a dedicated second layer of servers but can be obtained from information on server $W_i$. This is expressed by the conditions
\begin{equation} \label{eq:genb}
H(S^j_i | W_i ) = 0, ~~\text{for all $i,j$}.
\end{equation}
We add as conditions for the top layer
\begin{equation} \label{eq:WM}
H(M) = B, ~~H(W_j| M ) = 0, ~~\text{for all $j$}.
\end{equation}
The entropy conditions (\ref{en1})--(\ref{en3}) imply the conditions (\ref{eq:t1})--(\ref{eq:WM}) and thus the optimization problem for regenerating codes reduces to a problem of share sizes for a special version of a two-layer secret sharing scheme.

%This reflects that the encoding of $M$ into shares $W_j$ is detemrinistic. Any random component can be implemented as part of $M$, i.e. by giving u files size.

%We restrict ourselves to a configuration of $d+1$ servers even if the total number of servers $n > d+1$. Outer bounds obtained for $d+1$ servers will apply to any configuration containing $d+1$.
%\bigskip

%The optimization problem is, given $k$ and $d$, to charcterize the rate region, i.e. to characterize the admissable $B,\alpha,\beta,$ under the constraints
%(\ref{eq:t1}), (\ref{eq:t2}), (\ref{eq:b1}),  (\ref{eq:b2}),  \label{eq:genb}, \label{eq:gent}

%For a fixed linear ordering of the elements in $I$, let $S_I = ( S_i^{i'} : i > i' ).$ 

\section{Configurations of random variables} \label{S:2}

Let $k$ and $d$ be fixed. We make use of three different configurations of random variables. Minimal configurations appear in the upper bounds $B \leq B_q$, for $0 \leq q \leq k$. For a partition of $\{1,2,\ldots,d+1\}$ into subsets $V$, $M$ and $U$,with $|V|=q,$ $|M|=k-q,$ $|U|=d+1-k,$
\[
B \leq H( W_V, S_M, S_U^M) \leq B_q
\]
With the interpretation of the random variables as edges in an information-flow graph,  
\begin{equation} \label{S1}
S = \{ W_V, S_M, S_U^M \}
\end{equation}
corresponds to a min-cut \cite{DGWWR10}. 
For a partition of $\{1,2,\ldots,d+1\}$ into susbets $T,$ $L,$ $M$ and $U$, the two configurations
\begin{align}
&S =   \{ W_T, W_L, S_M^L, S_M, S_U^M \} \label{S3} \\
&S =  \{ W_T, W_L, S_M^L, S_U^L \}  \label{S2} 
\end{align}
will be used for improvements of the min-cut bounds. For $S$ as in (\ref{S3}),
\begin{equation} \label{R3}
H( S_M^L | W_T W_L S_M S_U^L ) = 0.
\end{equation}
For $S$ as in (\ref{S2}),
\begin{equation} \label{R2}
H(W_i | W_T W_{L \backslash i} S_M^i) \leq H(S_U^i) \qquad (i \in L)
\end{equation}

%The following summarizes our notation for the sizes in the three parititions of $\{1,2,\ldots,d+1\}$.
%\[
%\begin{array}{cclcl}
%(1) &~ &|V \cup M| = k, ~|V|=q, ~|M|=m,~|U|=d+1-k. \\
%(2,3) &~ &|T \cup L|=r, ~|T|=t, ~|L|=\ell,~|M|=m,~|U|=d+1-r-m. \\
%\end{array}
%\] 
We illustrate each of the three cases for a partition of $\{ 1, 2, \ldots, 8 \}$.
Putting a mark in position $i,i$ for $W_i \in S$ and a mark in position $i,j$ for $S_i^j \in S$, the
configurations are represented by the diagrams
\[
\begin{array}{ccccc}
\begin{array}{c|cccccc}
  &1 &2 &3 &4 &5 &6 \\ \hline
1 &\xa & & \\
2 & &\xa & \\
3 & & &-  \\
4 & & &\xc &- \\
5 & & &\xc &\xc &- \\ 
6 & & &\xc &\xc &\xc &- \\ 
7 & & &\xb &\xb &\xb &\xb \\
8 & & &\xb &\xb &\xb &\xb 
\end{array}
&\quad
&\begin{array}{c|ccccc}
  &1 &2 &3 &4 &5 \\ \hline
1 &\xa & & \\
2 & &\xt & \\
3 & & &\xt \\
4 & &\xd &\xd &- \\
5 & &\xd &\xd &\xc &- \\ 
6 & &    &    &\xb &\xb  \\ 
7 & &    &    &\xb &\xb  \\ 
8 & &    &    &\xb &\xb 
\end{array}
&\qquad
&\begin{array}{c|ccccc}
  &1 &2 &3 &4  \\ \hline
1 &\xa & & \\
2 & &\xa & & \\ 
3 & &  &\xt & \\ 
4 & &  &  &\xt  \\
5 & &  &\xd &\xd \\ 
6 & &  &\xd &\xd \\  
7 & &  &\xb &\xb \\ 
8 & &  &\xb &\xb 
\end{array} \\
{}\\[1.5ex]
(a) & &(b) & &(c) 
\end{array}
\]
\[
\begin{array}{cclcl}
(a) &~ &V = \{ 1,2 \}, ~M=\{ 3,4,5,6 \}, ~U=\{ 7,8 \} \\
(b) &~ &T = \{ 1 \}, ~L=\{ 2,3 \}, ~M=\{ 4,5\}, ~U=\{ 6,7,8 \} \\
(c) &~ &T = \{ 1,2 \}, ~L=\{ 3,4 \}, ~M=\{ 5,6 \}, ~U=\{ 7,8 \} 
\end{array}
\]
The goal in the next sections is to collect several copies of type (a), to break them into smaller pieces and to regroup them 
into a configuration of type (b) (Section \ref{S:4}), or type (c) (Section \ref{S:3}), or a combination of both (Section \ref{S:5}).

The proofs for the improved outer bound (\ref{new}) of a (4,3,3) code in Section \ref{S:2} are special cases. The reduction $H(W_2 S_3^2 S_4^3 S_1^3) = H(W_2 S_4^3 S_1^3)$ uses (\ref{R3}) with
$T = \emptyset, L = \{2\}, M= \{3\}, U = \{1,4\}$.  The reduction $H( W_2 | W_3 S_4^2) \leq H(S_1^2)$ uses (\ref{R2}) 
with $T = \emptyset, L = \{2, 3\}, M= \{4\}, U = \{1\}$.

\subsection{Parity check matrices}

Assume that $M$ is uniformly distributed on a vector space $V$ of dimension $B$, that the $W_i$ are uniformly distributed on vector spaces $V_i$ of dimension $\alpha$, and that the
$S_i^j$ are uniformly distributed on vector spaces $V_i^j$ of dimension $\beta$. We group the entropy conditions (\ref{en1})-(\ref{en3}) in a different way and give a vector space interpretation.
\begin{align}
&H(M) = B, ~~H(W_i) = \alpha,~~H(W_i|M) = 0 \label{C1} \\
&H(M|W_I) = 0~(|I|\geq k) \label{C2} \\
&H(W_i) = \alpha, ~~H(S_i^j) = \beta, ~~~H(S_i^j|W_i) =0 \label{C3} \\
&H(S_i^j|W_i) =0, ~~H(W_j|S_I^j) = 0~(|I| \geq d, j \not \in I). \label{C4}
\end{align}
Condition (\ref{C1}) implies that there exists a linear map $\phi_i : V \longrightarrow V_i$ of rank $\alpha$ with $phi_i(M) = W_i$. Together the maps define a linear encoder
\[
\phi = (\phi_1,\ldots,\phi_n) : V \longrightarrow \oplus V_i, ~~M \mapsto (W_1,\ldots,W_n).
\]
The image is a linear code of dimension $B$ and length $n \alpha$. 
Condition (\ref{C2}) implies that the generator matrix is of full rank $B$ on any submatrix of $k$ out of $n$ blocks of size $\alpha$.
Condition (\ref{C3}) implies that there exists a linear map of rank $\beta$
\[
\phi_i^j :  V_i \longrightarrow V_i^j, ~~~(W_i) \mapsto S_i^j.
\]
Condition (\ref{C4}) implies that there exists a linear map
\begin{equation}
\psi_I^j : \oplus V_i \longrightarrow V_j,~~~(W_i : i \in I) \mapsto W_j,~~~(|I| \geq d)
\end{equation}
and moreover that it factors as 
\begin{equation}
\psi_I^j : \oplus V_i \longrightarrow \oplus V_i^j \longrightarrow V_j,~~~(W_i : i \in I) \mapsto (S_i^j : i \in I) \mapsto W_j.
\end{equation}
The factorization allows us to characterize a regenerating code of length $n=d+1$ through the structure of its parity check matrix. 

\begin{proposition}
A linear code of dimension $B$ and length $n\alpha$ ($n$ consecutive blocks of length $\alpha$) represents a $(n,k,d)$ regenerating code with secondary 
parameters $(B,\alpha,\beta)$ if 
\[
\begin{array}{ll}
(1) &\text{Any $k$ blocks of size $\alpha$ have full rank $B$, and} \\
(2) &\text{Any $d+1$ blocks of size $\alpha$ satisfy parity checks of the form $H = H_{1\leq i,j \leq d+1}$, with} \\
      &\text{blocks $H_{i,j}$ of size $\alpha$ and of rank $\alpha$ on the diagional and of rank $\beta$ off the diagonal.} 
\end{array}
\]
\end{proposition}

For a code with $n=d+1$, the construction of a regenerating code is equivalent to the construction of a square block matrix $H$ of size $d+1$ with blocks of size $\alpha$ and with rank distribution 
\[
\left( \begin{array}{c|c|c|c}
\alpha &\beta  &\cdots &\beta \\ \hline
\beta  &\alpha &           &\beta \\ \hline
\vdots &           &\ddots &\vdots \\ \hline
\beta  &\beta   &\cdots &\alpha
\end{array} \right)
\]
such that columns in any $d-k+1$ blocks are independent (equivalent to (1)). For the special case $n=d+1,$ $k=d,$ the last condition is automatically fulfilled. Maximizing the rank $B$ is equivalent to minimizing the rank of the parity check matrix $H$.
 
\begin{example}
For a $(4,3,3)$ code with $\alpha=3, \beta=2$ we use 
\[
H = \left( \begin{array}{ccc|ccc|ccc|ccc}
1&0&0 &0&1&0 &0&0&1 &0&0&0 \\
0&1&0 &0&0&1 &0&0&0 &1&0&0 \\
0&0&1 &0&0&0 &1&0&0 &0&1&0 \\ \hline
0&0&0 &1&0&0 &0&1&0 &0&0&1 \\
1&0&0 &0&1&0 &0&0&1 &0&0&0 \\
0&1&0 &0&0&1 &0&0&0 &1&0&0 \\ \hline
0&0&1 &0&0&0 &1&0&0 &0&1&0 \\
0&0&0 &1&0&0 &0&1&0 &0&0&1 \\
1&0&0 &0&1&0 &0&0&1 &0&0&0 \\ \hline
0&1&0 &0&0&1 &0&0&0 &1&0&0  \\
0&0&1 &0&0&0 &1&0&0 &0&1&0 \\
0&0&0 &1&0&0 &0&1&0 &0&0&1
\end{array} \right)
\]
The $4 \times 4$ blocks are all equal to the identity matrix and thus the matrix has rank four and its row space is spanned by the first four rows.
The code is the one used inthe introduction.
\end{example}

\begin{proposition}
For any $d \geq 3$, there exists a code with$n=d+1, k=d, d=d$ and $B=(d-1)(d+1), \alpha=d, \beta=(d-1)$. Thus $B=B_{k-1} < B_k, B_{k-2}.$
\end{proposition}

We restrict to the case $n=d+1$. Outer bounds obtained for $(n,k,d)$ codes apply to $(n'>n,k,d)$ codes. 

\section{First outer bound} \label{S:4}

Let $S$ be a set of random variables, with each variable $X \in S$ of the form either $X = W_i$ or $X= S_i^j$. In the first case the entropy of $X$ is $H(W_i)=\alpha$ and in the second case it is $H(S_i^j) = \beta.$ The entropies $\alpha$ and $\beta$ serve as weights for the random variables in $S$ and the weight of $S$ is defined as $\sum_{X} H(X)$. By submodularity of the entropy function, the weight of $S$ is an upper bound for the entropy of $S$. 

\begin{proposition} \label{chain}
Let $\{ A_i : 1 \leq i \leq n \}$ and $\{ a_i : 1 \leq i \leq n \}$ be two sequences of sets of random variables such that for $1 \leq i < j \leq n$,
\begin{align}
H(M|A_i,a_i) = 0,  \\
H(a_j|A_i) = 0, 
\end{align}
Then
\[
n H(M) \leq \sum_{i=1}^n H(A_i) + H(a_1 \ldots a_n).
\]
In particular, for $H(a_n|a_1 \ldots a_{n-1}) = 0$,
\[
n H(M) \leq \sum_{i=1}^n H(A_i) +  \sum_{i=1}^{n-1} H(a_i).
\]
\begin{proof}
For $1 \leq i \leq n$,
\begin{align*}
H(M) &\leq H(A_i,a_i) \\
         &= H(A_i) + H(a_i|A_i) \\
         &\leq H(A_i) + H(a_i|a_{i+1} \cdots a_n) \\
\end{align*}
Finally, sum the $n$ inequalities and apply the chain rule.
\end{proof}
\end{proposition}

\begin{theorem} \label{thm-rs}
For given $k,d$, let $q,\ell,m$ be positive integers with $q+\ell+m \leq k$. Let $V_1,V_2,\ldots,V_{n-2}$ be subsets of $\{ \ell+m+1,\ldots,d+1 \}$ of size $q_i = |V_i| \leq k-\ell-m$ with empty intersection
$V_1 \cap V_2 \cap \cdots \cap V_{n-2} = \emptyset$. Then
\[
n B \leq B_q + \sum_{i=1}^{n-2} B_{q_i} + B_{\ell+m} - \ell m.
\]
\begin{proof}
For $d+1$ nodes $\{1,2,\ldots,d+1\}$, let $L = \{ 1,\ldots,\ell \}$, $M = \{ \ell+1,\ldots, \ell+m \}$, and denote by $U$ the set 
 $\{ \ell+m+1,\ldots,d+1 \}$ . Let $U_1 \cup U_2 \cup \cdots \cup U_{n-2}$ be a partition of $U$ such that, for each $i$, $U_i \cap V_i = \emptyset$. The empty intersection of the $V_i$ guarantees that such a partition exists. 
We apply the proposition with suitable choices for $\{ A_i : 1 \leq i \leq n \}$ and $\{ a_i : 1 \leq i \leq n \}$. Let 
\[
a_i = S_{U_i}^M ~~(1 \leq i \leq n-3)
\]
and let
\[
a_{n-2} = S_{U_{n-2}}^M \cup S_M, ~~a_{n-1} = W_L, ~~a_n = S_M^L.
\]
For each $a_i$ we choose $A_i$ such that $A_i \cup a_i$ is a minimal configuration and in particular $H(M|A_i,a_i)=0$. 
Recall from Section \ref{S:2} that a minimal configuration is determined by a partition $V' \cup M' \cup U'$. For $i=1,2,\ldots,n-2$, we choose 
\[
V'= V_i, ~~M' \supset L \cup M, ~~U' \supset U_i.
\]
For $i=n-1$, $V'= L \cup M$ and for $i=n$, $M' \supset L, U' \supset M$. With these choices $H(a_j|A_i) = 0$ for
$1 \leq i < j \leq n$. 
\end{proof}
\end{theorem}

\begin{corollary}
For $k=2p, d=3p$, $q=q_1=q_2=r+s=p$, it follows that $B \leq B_p - (p^2-1) \beta / 16$. This difference is unbounded as $p$ goes to infinity
\end{corollary}

Interpolation can be used to obtain similar estimates for other choices of parameters.

% In general, for a code with parameters $(n,k,d)$ we can fix any $k_0 < k $ nodes and any $d_0 < d+1-k$ nodes and assign fixed %roles to them: the $k_0$ nodes never fail and are always among the $k$ nodes chosen by a data collector, and the $d_0$ nodes %never fail and are always among the $d$ nodes assisting in repair of other nodes. The entropy $B$ of the original message is at %most $k_0 \alpha$ plus the capacity of a DDS with new parameters $k'=k-k_0$ and $d'=d-d_0-k_0.$ We use this to extend the %parameter space  for the bound in the theorem. 

\begin{example}
The case $(4,3,3)$.
Let $k=3, d=3$. For $q=r=s=1$ and for $V_1 = \{1\}, V_2 = \{4\}$, $4B \leq 3 B_1 + B_2 - \beta$ we list four minimal configurations and their division into $A \cup a.$
\[
\begin{array}{llllllllll}
        V' &M' &U'  &~~ &A &~ &a \\ \hline
   1  &2,3 &4  & &W_1, ~S_3^2,~S_4^2 & &S_4^3  \\
  4  &2,3 &1  & &W_4,  ~S_3^2,~S_1^2 & &S_1^3  \\
   2,3  &4 &1  & &W_3,  ~S_4^1              & &W_2   \\
1  &2,3 &4  & &W_1,  ~S_4^2,~S_4^3 & &S_3^2 
\end{array}
%\qquad
%\begin{array}{c|ccccc}
%     &2    &3  \\ \hline
%2   &\xt & \\
%3   &\xd &- \\
%1   &      &\xb \\
%4   &      &\xb
%\end{array} \\
\]
The fourth row is used as an upper bound for $H(S_3^2)$ but we can avoid it and use $H(S_3^2) = \beta$. Then the bound becomes  $3B \leq 2 B_1 + B_2 - \beta.$ This improvement applies whenever the theorem is used with $\ell = m = 1.$
\end{example}

\begin{proposition} \label{P1}
For $2 \leq p \leq k-2$, 
\[
3B \leq 2 B_p + B_{p\pm1} - \beta 
\]
\begin{proof}
For both the plus sign and minus sign we partition $d+1$ nodes into $T \cup \ell \cup m \cup V \cup U$ and we fix an ordering on each of $T$, $V$ and $U$. For the plus sign we choose the sets of size $|T|=p-1$, $|V|=k-p-1 \geq 1$ and $|U|=d+1-k \geq 1$. Let $v \in V$ be the last element in $V$ and let $u \in U$ be the last element in $U$.
We apply Proposition $\ref{chain}$ with three minimal configurations. For each, we list $V'$, $M'$, $U'$ and $a$.
\[
\begin{array}{lclclclc}
V' & &M' & &U'  & &a \\ \hline
T,v  &~ &\ell, m, V \backslash v &~ &U &~~ &S_U^m \\
T,u &~  &\ell, m, V \backslash v &~ &v, U\backslash u &~~ &S_V^m, S_V \\
T,\ell,m &~ &V                           &~ &U &~~ &W_T, W_{\ell}
\end{array}
\]
For the minus sign we choose the sets of size $|T|=p-2$, $|V|=k-p \geq 2$ and $|U|=d+1-k \geq 1$.
Let $v, v' \in V$ be the last elements in $V$ and let $u \in U$ be the last element in $U$.
We apply Proposition $\ref{chain}$ with three minimal configurations. For each, we list $V'$, $M'$, $U'$ and $a_i$.
\[
\begin{array}{lclclclc}
V' & &M' & &U'  & &a \\ \hline
T,v,v'  &~ &\ell, m, V \backslash \{ v,v' \} &~ &U &~~ &S_U^m \\
T,u &~  &\ell, m, V \backslash v' &~ &v', U\backslash u &~~ &S_V^m, S_V \\
T,\ell,m &~ &V                           &~ &U &~~ &W_T, W_{\ell}
\end{array}
\]
\end{proof}
\end{proposition}

\begin{example} \label{E1}
For a $(8,6,7)$ code we apply the theorem with $q=r=s=2$, $V_1 = \{ 5,6 \}, V_2 = \{ 7,8 \}$. Then
\[
4 B \leq 3 B_2 + B_4 - 2 \cdot 2 \beta = 2 B_2 +  2 B_3 - 3 \beta = 10 \alpha + 43 \beta.
\]
This is less than the functional repair outer bound in the range $23 / 6 < \alpha / \beta < 37 / 6$.  The gap reaches a maximum at $\alpha= 5 \beta$  where it lowers the bound $B \leq 24 \beta $ by $3 \beta / 4$. We will compare this with other bounds in the next section.
\end{example}
%In the next section we will see different gaps $4/7 < 7 / 12 < 2/ 3 < 3/ 4$.

\section{Second outer bound} \label{S:3}

In graph terms, we consider the complete graph on $d+1$ vertices, with edges $\{ W_i \}$ and $\{ S_i^j \}$. The variable $S_i^j$ connects node $i$ and node $j$. The variable $W_i$ connects node $i$ with node $i$ or, after creating two copies of node $i$, node $i$-in with node $i$-out. The sets $S$ define subgraphs with a block structure on the adjacency matrix. Connections between nodes can be interpreted as channels of bandwith $H(W_i)=\alpha$ or $H(S_i^j)=\beta$.

For a regenerating code with parameters $(n,k,d)$ the repair matrix for $d+1$ nodes is a square matrix of size $d+1$ that indicates which nodes function properly (a nonzero entry in the position $i,i$ for node $i$) and which nodes help other nodes (a nonzero entry in position $i,j$ for node $i$ helping node $j$) \cite{}. A minimal configuration has $q$ diagonal entries and $k-q$ nonzero columns. For $d=7, k=6, q=2,$
\[
\begin{array}{cccccc}
\ast \\
& \ast \\
& &- \\
& &\xb &- \\
& &\xb &\xb &- \\
& &\xb &\xb &\xb &- \\ 
& &\xb &\xb &\xb &\xb \\ 
& &\xb &\xb &\xb &\xb
\end{array}
\]
By replacing $\ast$ with $\alpha$ and $\xb$ with $\beta$, the configuration corresponds one-to-one to the adjacency matrix of an acyclic graph on $d+1$ vertices, with loops of weight $\alpha$ and other edges of weight $\beta$. It is common in this setting to think of a loop as an edge between two copies of the same node, an input node and an output node \cite{DGWWR10}. The total weight of the edges, or the weight of the configuration, is given by
\[
B_q = q\alpha +  \binom{k-q}{2} \beta + (d+1-k)(k-q) \beta.
\]
The information stored in node $i$ is modeled by the random variable $W_i$ and the helper information from node $i$ to node $j$ by the random variable $S_i^j$. The entropy of the random variables is bounded by $H(W_i) \leq \alpha$ and $H(S_i^j) \leq \beta$. The entropy of the message that a destination can recover from the configuration of nodes is bounded by the weight of the configuration. The minimal configurations give $H(M) \leq \min_q B_q$, the minimum taken over all $q \in \{0,1,\ldots,k\}$.

\bigskip

As observed in \cite{SRKR12a} and developed in \cite{SSK13}, variables $S_m^\ell$, $\ell \in L$,  in row $m$ of the configuration, all correspond to information from node $m$, i.e to information from a source of bounded entropy $H(W_i) \leq \alpha$. For $L$ large enough, some nonzero mutual information among the $S_m^\ell$, $\ell \in L$, is expected. This is captured by Proposition 2 in \cite{SSK13}. Equation (18) in the next to last line of the proof of Proposition 2 in \cite{SSK13} states that the sum of the entropies $H(W_R|W_m)$, $H(W_R)$ ($\ell-1$ times) and
$H(S^L_m),$ is stricly less than the trivial upper bound $\ell (r \alpha + \beta)$ whenever $\alpha > (d-r) \beta$. We use the claim with the singleton $m$ replaced by a set $M$ and adapt the proof.

\begin{proposition}[Equation (18) in Proposition 2 \cite{SSK13}] \label{P0}
Let $ L \subset R$, $R \cap M = \emptyset$, $|L|=\ell, |M|=m, |R|=r,$ $r+m \leq k$.
\[
H(W_R|W_M) + (\ell-1) H(W_R) + H(S^L_M) \leq \ell ( r \alpha + \beta) + \ell( (d-r)\beta-\alpha )
\]
\begin{proof}
Combine Equations (\ref{p21}) and (\ref{p22}).
\begin{align}
     H(W_R|W_M) + (\ell-1)H(W_R) + H(S^L_M)  \nonumber
\leq~ &H(W_R|S^L_M) + (\ell-1) H(W_R) + H(S^L_M) \nonumber \\
                       =~ &H(S^L_M|W_R) + H(W_R) + (\ell-1) H(W_R) \nonumber \\
                       \leq~ &\sum_{i \in L} ( H(S^i_M|W_R) + H(W_R) ). \label{p21}
\end{align}
For each $i \in L$,
\begin{align} H(S^i_M|W_R) + H(W_R)   \nonumber
                       =~ &H(W_i | W_{R\backslash i} S^i_M ) + H (W_{R\backslash i} S^i_M)   \nonumber \\
                       \leq~ &H(W_i | W_{R\backslash i} S^i_M ) + 
                                H (W_{R\backslash i})  + H(S^i_M)   \nonumber \\
                       \leq~ &(d-r+1-m)\beta + (r-1) \alpha  + m \beta  \nonumber \\
                       =~ &(d-r)\beta-\alpha  +  r \alpha + \beta   \label{p22} 
%                       =~ &B_{r}-B_{r+1} + r \alpha + \beta. 
\end{align}
\end{proof}
\end{proposition}

Turning the proposition into an improved outer bound follows a standard procedure. In \cite{SSK13}, the improvement is applied, for given $\alpha$ and $\beta$, to the outer bound $B_p$ that is minimal, for the given $\alpha$ and $\beta$, among all $B_q$, $q=0,1,\ldots,k$.
This leaves open the possibility that the best overall outer bound for given $\alpha$ and $\beta$ comes from improving a $B_q$ different from $B_p$. For that reason we change the order and first collect a sequence of improved upper bounds and then address later which upper bound is optimal for which choice of $\alpha$ and $\beta$. Other differences will be pointed out after the statement of the theorem and illustrated by an example.

\bigskip

In the next thereom, the minimal configuration with $q$ intact nodes and $k-q$ nodes being repaired refers to the configuration at the beginning of the section. The region of helper information $S_i^j$ is bounded by
$q+1 \leq j \leq k$, $j < i \leq d+1$.
 
\begin{theorem} \label{thm-lm}
Let $\cup_{(M,L)} S_M^L$ be a disjoint union of the helper information in a minimal configuration with $q$ intact nodes and $k-q$ nodes being repaired.
For each $(M,L)$, let $\ell=|L|$, $|M|=m$, $r \geq \ell$. Then
\[ 
B + \sum_{(M,L)} \ell B  \leq B_q + \sum_{(M,L)} ( B_{r+m-1} + (\ell-1)(B_{r+m-2}-\beta) ). 
% B_p + \sum_{(m,L)} (\ell-1) B_s + \sum_{(m,L)} B_t + \sum_{(m,L)} \ell(B_{r}-B_{r+1}) 
\]
%For $(d-p) \leq \bar \alpha \leq (d-p+1)$, the bound reaches its optimum for
%$q \in \{p-1,p\}$ and 
%$r \in \{ \max(\ell,p), p+1 \}$.
\begin{proof}
Let $\sum_{(M,L)} H(S_M^L)$ be a disjoint sum of the $S$ contributions to $B_q$.
%, i.e. of the trapezium with $k-q$ columns of decreasing  height $(d-p), \ldots, (d+1-k)$. Then
%Starting with the three outer bounds 
\begin{equation}
B \leq B_q + \sum_{(M,L)} ( H(S_M^L) - \ell m \beta )  \qquad (\ell \leq k-q) \label{eq1} 
\end{equation}
Fix a term $(M,L)$  and let $L \subseteq R$, $r = |R|$.
\begin{align}
B &\leq B_s + H(W_R) - r \alpha \qquad (s \geq r) \label{eq2}, \\
B &\leq B_t + H(W_R|W_M) + H(W_M) - (r+m)\alpha \qquad (t \geq r+m) \label{eq3}
\end{align}
\begin{align*}
B + \sum_{(M,L)} \ell B  &\leq  %B_q + \sum_{(M,L)} (2 \ell-1) B_r - \sum_{(m,L)} (\ell-1) B_{r+1} \\ 
                          B_q + \sum_{(M,L)} ( B_{r+m-1} + (\ell-1)(B_{r+m-2}-\beta) ). 
\end{align*}
\end{proof}
\end{theorem}

\begin{example} \label{E544}
For the $(5,4,4)$ code, Theorem \ref{thm-lm} yields the following upper bounds. Each upper bound is obtained as the average of the linear forms above it. Below each upper bound is the range for $\bar \alpha = \alpha / \beta$ where it is minimal among 
the given six upper bounds. 
\[
\begin{array}{lclcl}
\begin{array}{lll}
{} \\
B_4 &{} \\[1.5ex]
\end{array}
&\quad
&\begin{array}{lll}
{} \\
B_3 &{} \\[1.5ex]
\end{array}
&\quad 
&\begin{array}{lll}
B_2 \\
B_3 &B_2^- \\[1.5ex]
\end{array} \\
\bar B \leq 4 \bar \alpha & &\bar B \leq 3 \bar \alpha + 1 & &\bar B \leq (7 \bar \alpha + 6) / 3 \\[1.5ex]
{\bar \alpha \in [0,1]} & &\bar \alpha \in [1,3/2] & &\bar \alpha \in [3/2,2] \\[3ex]
\begin{array}{lll}
B_1 \\
B_3 &B_2^{-} \\
B_3 &B_2^{-} &B_2^{-} \\[1.5ex]
\end{array}
&\quad 
&\begin{array}{lll}
B_1 \\
B_2 &B_1^{-} \\
B_3 &B_2^{-} &B_2^{-} \\[1.5ex]
\end{array}
&\quad 
&\begin{array}{lll}
B_1 \\
B_2 &B_1^{-} \\
B_2 &B_1^{-} &- \\[1.5ex]
\end{array} \\ 
\bar B \leq (13 \bar \alpha + 14 ) / 6  & &\bar B \leq (11 \bar \alpha + 19 ) / 6   & &\bar B \leq (7 \bar \alpha + 22 ) / 5 \\[1.5ex]
\bar \alpha \in [2,5/2] & &\bar \alpha \in [5/2,37/13] & &\bar \alpha \in [37/13,4] 
\end{array}
\]
\end{example}

 The functional repair outer bound $B \leq \min B_q$ attains its minimum in $B \leq B_p$ on the interval $(d-p)\beta \leq \alpha \leq (d-p+1)\beta$. We give a short proof that the exact repair outer bound is strictly less than the functional repair outer bound for all $(d-k+3/2)\beta < \alpha < d \beta$. Let $\hat B = \min B_q$. We use Theorem \ref{thm-lm} to give a different proof of Proposition \ref{P1}. 

\begin{proposition} \label{P2}
For $2 \leq p \leq k-2$, 
\[
3B \leq 2 B_p + B_{p\pm1} - \beta 
\]
Moreover,
\begin{align*}
&(1)~~2 B_p + B_{p+1} - \beta \leq 3 B_p - \beta/2,    \quad \text{for $(d-p) \leq \alpha \leq (d-p+1/2)\beta.$} \\
&(2)~~2 B_p + B_{p-1} - \beta \leq 3 B_p - \beta/2,  \quad \text{for $(d-p+1/2) \leq \alpha \leq (d-p+1)\beta.$}
\end{align*}
\begin{proof}
For $2 \leq p \leq k-2$, we apply the proposition with $q=p$, $\ell =2, r= p$ :
\[
3B \leq B_p + B_p + B_{p-1} - \beta
\]
or with $q=p$, $\ell =2, r= p+1$ :
\[
3B \leq B_p + B_{p+1} + B_p - \beta
\]
Inequality (1) follows from $B_{p+1}-B_p = \alpha - (d-p)\beta$ and Inequality (2) similarly from
$ B_{p-1}- B_p = (d-p+1)\beta - \alpha$. Together they proof $3B \leq 3 \hat B - \beta/2$ on the interval $(d-p) \leq \alpha \leq (d-p+1)\beta.$ 
Now take the union of the intervals for $2 \leq p \leq k-2$.
\end{proof}
\end{proposition}

The next Corollary follows from either Proposition \ref{P1} in the previous section or from Proposition \ref{P2} above. Thus, methods in either section can be used to prove a non-vanishing gap between the functional-repair and exact-repair outer bounds.

\begin{corollary}
As a result, the exact repair capacity $B$ of an $(n,k,d;B,\alpha,\beta)$ regenerating code satsfies
$B \leq \hat B- \beta/ 6$ on the interval $(d-k+2) \beta \leq \alpha \leq (d-1) \beta$.
\end{corollary}

\section{Using the bounds in combination} \label{S:5}

Theorem \ref{thm-rs} and Theorem \ref{thm-lm} give two different outer bounds for exact-repair regenerating codes. Both are linear and are valid in the full range of $\alpha / \beta$. Using the bounds with different choices for the parameters and then taking the minimum over all choices will produce a piece-wise linear upper bound for $B$. We provide some details on using the bounds in combination for the cases $(n,k,d)=(8,6,7)$ and $(n,k,d)=(5,4,4)$.
 
Figure 1 gives outer bounds for a $(8,6,7)$ code. They are presented as a trade-off between $\alpha/B$ and $\beta/B$ and are based on Theorem \ref{thm-lm}. In this case, the graph for the trade-off shows the differences more clearly than the upper bound graph with $B / \beta$ as a function of $\alpha / \beta$.

The lowest of the four outer bounds is the functional-repair outer bound $B \leq \min_q ( B_q : 0 \leq q \leq k ).$ Next we apply the theorem with the choices of \cite[Proposition 2 and Theorem 1]{SSK13}, i.e. $M = \{ m \}$ is a singleton and $\ell$ is fixed.
Next we allow choices with different $\ell$ and finally we allow $M$ of various sizes. 

\begin{example}
We illustrate the improvements for $\alpha = 5 \beta.$ This corresponds to the vertex $B_3 = B_2$ in the functional-repair outer bound: $B \leq B_3 = 3 \alpha + 9 \beta$, $B \leq B_2 = 2 \alpha + 14 \beta.$ The various improvements are the following. In each case we use Theorem \ref{thm-lm}.

For $q=3$ and two choices with $\ell=3$, as in \cite[Proposition 2 and Theorem 1]{SSK13} :
\[
7 B \leq B_3 + 2 ( B_3 + B_2 + B_2 - 2\beta) =  4 B_2 + 3B_3 - 4.
\]
For $q=2$ and choices $\ell=2$ ($1 \times$) and $\ell=3$ ($3 \times$):
\[
12 B \leq B_2 + (B_3 + B_2 - \beta) + 3 ( B_3 + B_2 + B_2 - 2 \beta )  = 8 B_2 + 4 B_3 - 7.
\]
For a choice with $m > 1$, let $q=2$ and $\ell=2, m=2$ ($1 \times$) and $\ell=3, m=2$ ($1 \times$):
\[
6 B \leq B_2 + (B_4 + B_2 - 2 \beta) + (B_4 + B_2 + B_2 - 4 \beta) = 2 B_2 + 4 B_3 - 4.
\]
or
\[
6 B \leq B_2 + (B_3 + B_1 - 2\beta) + (B_4 + B_2 + B_2 - 4\beta) = 4 B_2 + 2 B_3 - 4.
\]
At $B_2 = B_3$, the gaps with the fractional-repair upper bound increase as $4/7 < 7/12 < 2/3$. But none reaches the gap of $3/4$ that was found in Example \ref{E1}, using Theorem \ref{thm-rs}:
\[
4 B \leq 2 B_2 +  2 B_3 - 3.
\]
\end{example}

\begin{example}
Example \ref{E544} lists upper bounds for a $(5,4,4)$ code that are obtained with Theorem \ref{thm-lm}. In particular,
\begin{equation} \label{B5}
5B \leq 7  \alpha + 22 \beta = 2 B_2 + 3 B_1 - 2 \beta.
\end{equation}
Here $B \leq B_1 = \alpha + 6 \beta$ and $B \leq B_2 = 2 \alpha + 3 \beta$. Theorem \ref{thm-rs} applied with 
$r+s=3$ and $q=q_1=q_2=1$ gives
\[
4 B \leq 3 B_1 + B_3 - 2\beta  = 2 B_1 + 2 B_2 - \beta,
\]
which yields no improvement to the bounds listed in Example \ref{E544}. However, after exploiting nonzero mutual information among the $a_i$, the theorem uses trivial estimates for each of the $H(A_i)$. In this case, together the $A_i$ contain five copies of $S_m^L$, each with $\ell=2$. Using Proposition \ref{P0} we can therefore improve the last bound to
\[
4B + 5 \cdot 2B \leq  2 B_1 + 2 B_2 - \beta + 5 \cdot (B_1 + B_2 - \beta)
\]
0r 
\begin{equation} \label{B14}
14 B \leq 7 B_1 + 7 B_2 - 6 \beta.
\end{equation}
Comparing (\ref{B14}) and (\ref{B5}) at $\alpha = 3 \beta$ ($B_1 = B_2$), $B \leq B_1 - 3/7 < B_1 - 2/5.$ And (\ref{B14}) improves on the bounds in Example \ref{E544} for $19/7 < \alpha/\beta < 23/7.$ 
\end{example}

%\nocite{RSK11},\nocite{T14}, \nocite{SSK13}, \nocite{DRWS11}, \nocite{SRKR12a}, \nocite{SRKR12b}, \nocite{SRKR10},  %\nocite{DGWWR10}, \nocite{GPV13}, \nocite{GRCP13}

% \cite{RSKR09}, \cite{SRK11}, 

\newpage

%\bibliographystyle{plain}
%\bibliography{regen}

\newpage

\begin{figure}[!htb]
\centering
\includegraphics[scale=.8]{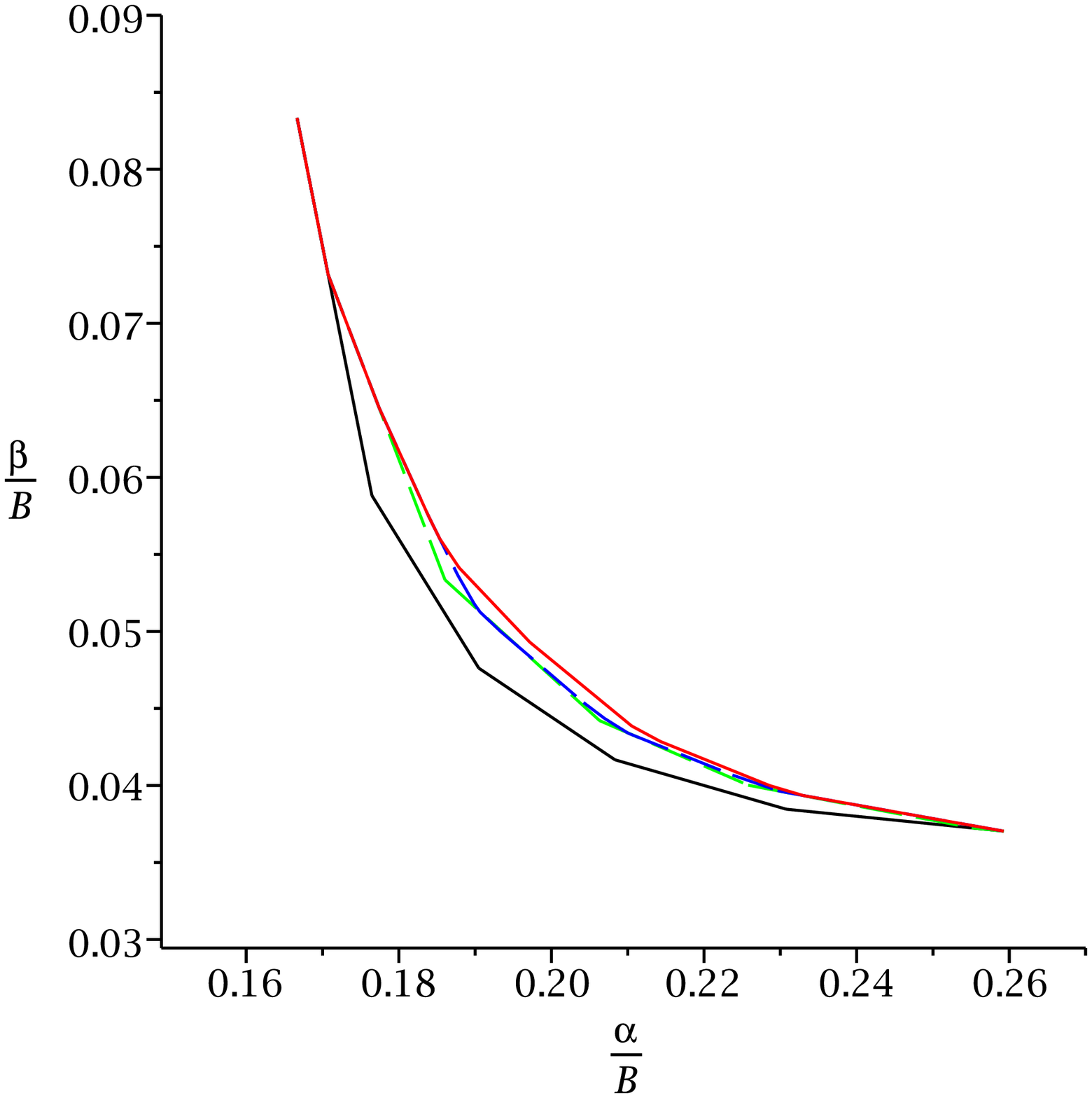}
\caption{Outer bounds for $(n,k,d)=(8,6,7)$}
\label{fig:867}
\end{figure}

\end{document}